\documentclass[conference]{IEEEtran}
\IEEEoverridecommandlockouts
\usepackage{subfig}
\usepackage{multirow}
\usepackage{cite}
\usepackage{amsmath,amssymb,amsfonts}
\usepackage{algorithm}
\usepackage[noend]{algpseudocode}
\usepackage{graphicx}
\usepackage{textcomp}
\usepackage{xcolor}
\usepackage{multirow}
\usepackage{paralist}
\usepackage[font=footnotesize]{caption}

\graphicspath{{./Images/}}

\def\BibTeX{{\rm B\kern-.05em{\sc i\kern-.025em b}\kern-.08em
    T\kern-.1667em\lower.7ex\hbox{E}\kern-.125emX}}
\begin{document}

\title{Examining Untempered Social Media: \\Analyzing Cascades of Polarized Conversations}

\author{\IEEEauthorblockN{Arunkumar Bagavathi\IEEEauthorrefmark{1}, Pedram Bashiri\IEEEauthorrefmark{1}, Shannon Reid\IEEEauthorrefmark{2}, Matthew Phillips\IEEEauthorrefmark{2}, Siddharth Krishnan\IEEEauthorrefmark{1}}
\IEEEauthorblockA{\textit{\IEEEauthorrefmark{1}Dept. of Computer Science, \IEEEauthorrefmark{2} Dept. of Criminal Justice} \\
\textit{University of North Carolina at Charlotte}\\
abagavat@uncc.edu, pbashiri@uncc.edu, sreid33@uncc.edu, mdphill1@uncc.edu, skrishnan@uncc.edu,}
}

\maketitle

\begin{abstract}
Online social media, periodically serves as a platform for cascading polarizing topics of conversation. The inherent community structure present in online social networks (homophily) and the advent of fringe outlets like Gab have created online ``echo chambers" that amplify the effects of polarization, which fuels detrimental behavior. Recently, in October 2018, Gab made headlines when it was revealed that Robert Bowers, the individual behind the Pittsburgh Synagogue massacre, was an active member of this social media site and used it to express his anti-Semitic views and discuss conspiracy theories. Thus to address the need of automated data-driven analyses of such fringe outlets, this research proposes novel methods to discover topics that are prevalent in Gab and how they cascade within the network. Specifically, using approximately 34 million posts, and 3.7 million cascading conversation threads with close to 300k users; we demonstrate that there are essentially five cascading patterns that manifest in Gab and the most ``viral" ones begin with an echo-chamber pattern and grow out to the entire network. Also, we empirically show, through two models viz. \emph{Susceptible-Infected} and \emph{Bass}, how the cascades structurally evolve from one of the five patterns to the other based on the topic of the conversation with upto 84\% accuracy.

\end{abstract}


\section{Introduction}
Fringe social media sites, such as Gab, 8chan, and PewTube, have become a fertile ground for individuals and groups with far right and extreme far right views to post and share their messages in an unfettered manner and to galvanize supporters for their cause~\cite{zannettou2018gab}. While most mainstream social media like Reddit, Twitter, and Facebook moderate their content and deplatform more extreme users and groups, the emergence of outlets like 8chan and Gab.ai have given radical groups large content delivery networks to broadcast their polarizing messages. These social networks have morphed into alt-right echo chambers~\cite{lima2018inside,garimella2018political} and have garnered close to 450,000 users. The echo chamber effect is often amplified via online conversations and interactions that occur on social networks.  In recent times, we have observed that such interactions, combined with the exploitation of online social networks~\cite{Moreno2011Dynamics} is an effective strategy to recruit members, instigate the public, and ultimately culminate in riots and violence as witnessed recently in Charlottesville and Portland. Due to the threat of violence that these groups bring with them, analyzing the online dynamics of conversations and interactions in such social media sites is an important problem that the research community is trying to address~\cite{kim2012you,mathew2018spread}.


In this work, using the well-known propagation mechanism of \emph{information cascades}~\cite{cheng2014can,krishnan2016seeing}, we demonstrate how polarized conversations take shape on Gab. By analyzing \textit{34M} posts and \textit{3.7M} cascades built from conversations on Gab, we show that there are five different classes of conversation cascades, where each type shows varied level of user participation, and responses. By analyzing the types of cascades, we give post-level intuition and several structural properties of these cascades. To emphasize the post level details of the cascades, we present an algorithm to classify hashtags, and eventually cascades into topics. We observe that controversial topics are adopted by users that are more strongly connected and also these topics generate larger cascades. By analyzing structural dynamics of conversation cascades on Gab, we observe that all cascades start with a simple linear pattern and evolve into other patterns when a topic becomes viral and more users join the conversation. We found that the average time and average posts to make an evolution on Gab is 1.5 days and atleast 3 posts respectively on average. We model this evolution of cascades using popular network growth models like \textit{Susceptible-Infected} model~\cite{banks2013growth} and \textit{Bass} model~\cite{mahajan1990new}. Our best model fit give upto 84\% accuracy.

Essentially, we answer the following research questions with our corresponding contributions in this broad study of Gab:

\begin{itemize}
	\item \textbf{What are the types of cascading behavior in Gab conversations?} We study conversation patterns in Gab as cascades and provide metrics to measure structural patterns within conversations. Our analysis show that the rarely occurring cascades get viral even when their user response rates are much lower compared to commonly occurring cascades.
	
	\item \textbf{Can we characterize user relationship and topics of Gab conversations?} We propose an algorithm to cluster topics that circulate in conversation cascades and our results show that the polarizing topics gain lot of traction in Gab conversations.
	\item \textbf{Can we study evolution of Gab conversations using the cascades?} With Gab conversations represented as cascades, we give pathways for these cascades to evolve over time. To capture this evolution patterns algorithmically, we use the \textit{Susceptible-Infected} model~\cite{banks2013growth} and the \textit{Bass} model~\cite{mahajan1990new}. We show applicability of different models to different evolution types.
	


\end{itemize}

\section{Related Work}
Related work of our research can be split into three sections: \begin{inparaenum}[i)]
	\item Applications of Gab analysis
	\item Cascading behavior in social media
	\item Quantitative analysis of topics in social media
\end{inparaenum}.

The use of social media to study radicalization~\cite{oro48477}, discrimination~\cite{ottoni2018analyzing}, and fringe web communities~\cite{zannettou2018origins} is gaining traction over the past few years. Particularly, past studies highlights the fact that the advent of Gab.com creates a scope to analyze topics like alt-right echo-chamber and hate speech research~\cite{zannettou2018gab,lima2018inside}. Most notably, the recent work studied the spread of hate speech among Gab users with the help of repost cascades and friend/follower network~\cite{mathew2018spread}. Our research adds an extra dimension to existing Gab works, in which we analyze its conversations, provide analysis on growth of topics, and give measures and metrics to study conversation structure and evolution in the perspective of cascades.


Cascades in social media accounts for information dissemination~\cite{cheng2014can}, which can be applied to variety of applications like fake news~\cite{vosoughi2018spread}, viral marketing~\cite{chang2015persuasive}, and emergency management~\cite{kim2018social}. These information cascades are crucial in incorporating machine learning models for variety of applications like: modeling influence propagation in the social media~\cite{matsubara2012rise}, predicting number of reshares using self-exciting point process~\cite{zhao2015seismic}, and modeling network growth patterns as an alternative to sigmoid models~\cite{zang2016beyond}. In this work, we reframe information cascades as conversation cascades and give novel ideas on defining models for cascade evolution types in Gab conversations. 


Just like Twitter~\cite{Wang:2011:TSA:2063576.2063726}, hashtags on Gab are means to add metadata to posts that highlights topics. Some researchers defined the topic (class) of hashtags manually~\cite{10.1007/978-3-319-11538-2_30}~\cite{quantitaveApproachToAntisemitism}. Lee et. al.~\cite{6137387} classify Twitter Trending Topics into 18 general categories using one bag-of-words text based approach and one network approach. Wang et. al~\cite{7412726} propose Hashtag Graphbased Topic Model (HGTM) to discover topics of tweets. We believe manual labeling is not scalable and unsupervised algorithms are not accurate enough due to the nature of hashtags, thus we propose a supervised semi-automated procedure to classify hashtags.



\section{Preliminaries}
In this section, we briefly describe about the dataset that we use in all experiments, terminologies in our methods and statistics of cascades and topics in the dataset.
\subsection{Dataset Description}
\begin{figure}
\centering
\includegraphics[scale=0.5]{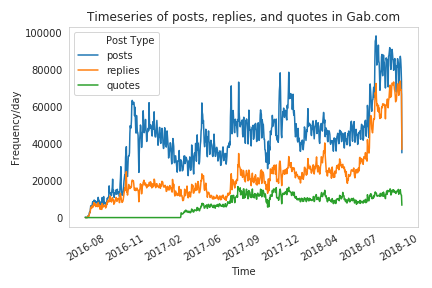}
\caption{Timeseries of frequency posts, replies, and reshares from the origin of gab.com(August 2016) until the forum went down on the last week of October 2018} \label{fig:post_timeseries}
\end{figure}

Gab.com/Gab.ai is a social media forum, founded in 2016, provide a forum to connect and share information among users. Even though the description of the forum looks very similar to most popular social media counterparts like Twitter and Facebook, Gab is known to support individual liberty and committed to contribute for free speech in the social media community\footnote{https://gab.com/}. However, Gab has strong restriction policies on posts and users, which are promoting pornography, terrorism and violence. Users of Gab can share information via \textit{posts}, \textit{post replies}, and \textit{quotes/reshares}. We use the Gab data published by data scraping forum \textit{pushshift.io}\footnote{https://pushshift.io}. Figure~\ref{fig:post_timeseries} gives an overview of the dataset as timeseries plot for the number of posts, replies, and quotes appeared in Gab between August 2016 and October 2018.

The dataset is a comprehensive collection with \textbf{34 million} posts, replies, quotes posted  between the date range of August 2016 and October 2018, about \textbf{15,000} groups and about \textbf{300,000} public users information. It is evidential from Figure~\ref{fig:post_timeseries} that our dataset comprise of 55\% posts, 30\% replies, and 15\% quotes. This data is available with complete set of metadata like \textit{time, attachments, likes, dislikes, replies, quotes} along with post and user details. 



\subsection{Conversation cascades}
\label{sect:cascades}
Microblogging conversations have been widely studied in the context of cascades for a wide spectrum of applications like emotion analysis~\cite{kim2012you}, topic modelling~\cite{alvarez2016topic} and cascade analysis~\cite{choi2015characterizing}. In this work, we give variety of cascade representations for conversations in Gab and give their in-depth structural and temporal analysis, response rates, and longevity.

\begin{figure}[!ht]
	\centering	
		\subfloat[Figure a: Cascade types\label{fig:cascade_type}]{\includegraphics[scale=0.5]{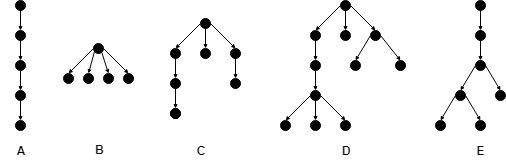}}	
	
		\qquad	
	
		\subfloat[Figure b: No. of posts per cascade type\label{fig:cascade_freq}]{\includegraphics[scale=0.5]{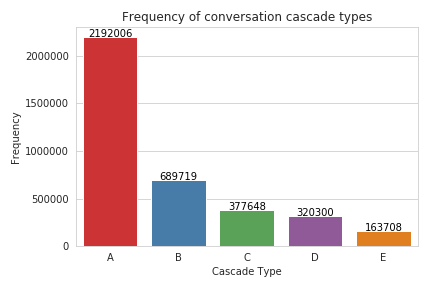}}

		\caption{Possible shallow cascading structures in replies and reshares of posts and number of posts in each cascade type. As predicted, most of the cascades follow simple patterns(type \textit{A}). Interestingly, many conversations form cascade type \textit{B} in which many response come directly to the root post and does not evolve into any other cascade types. Small number of conversations follow cascade type \textit{C}, which split from the root and each branch from root follows linear pattern. Very few conversations follow cascade type \textit{D} pattern, which has highly nested structure, both at root level and branch level and cascade type \textit{E}, which initially follows linear pattern and takes non-linear during the evolution time.}
		\label{fig:cascadetype}
\end{figure}



\begin{table*}[!t]
\caption{Basic statistics of depth, volume, number of unique users, and structural virality(Wiener Index) of all cascade types. Overall, the cascade type E achieves higher structural popularity even though the number of unique user participation is low.} \centering
\label{tab:cascade_statistics}
\centering

\begin{tabular}{|c|p{0.4cm}|c|p{0.4cm}|c|p{0.4cm}|c|p{0.4cm}|c|p{0.4cm}|c|p{0.4cm}|c|c|c|p{0.4cm}|c|}
	\hline
	\textbf{Metric} & \multicolumn{4}{c|}{\textbf{Depth}} & \multicolumn{4}{c|}{\textbf{Volume}} & \multicolumn{4}{c|}{\textbf{Users}} & \multicolumn{4}{c|}{\textbf{Wiener Index}} \\
	\hline
	 & Min. & Max. & Avg. & Std. Dev. & Min. & Max. & Avg. & Std. Dev. & Min. & Max. & Avg. & Std. Dev. & Min. & Max. & Avg. & Std. Dev. \\
	\hline
	A & 1 & 261 & 0.91 & 2.52 & 2 & 279 & 2.33 & 1.09 & 2 & 110 & 1.89 & 0.41 & 0.06 & 43.67 & 0.57 & 0.21\\
	\hline
	B & 1 & 2 & NA & NA & 3 & 209 & 4.1 & 2.38 & 3 & 200 & 2.27 & 6.41 & 0.004 & 0.33 & 0.38 & 0.22\\
	\hline
	C & 2 & 64 & 2.79 & 1.07 & 4 & 191 & 8.59 & 3.25 & 4 & 175 & 5.27 & 16.19 & 0.005 & 8.5 & 0.37 & 0.22\\
	\hline
	D & 2 & 209 & 7.81 & 2.54 & 5 & 3748 & 24.53 & 16.45 & 5 & 1109 & 11.0 & 16.19 & 0.002 & 22.53 & 0.45 & 0.47\\
	\hline
	E & 2 & 324 & 6.9 & 4.9 & 4 & 1629 & 15.72 & 10.76 & 4 & 280 & 3.17 & 2.88 & 0.01 & 38.07 & 0.93 & 0.7\\
	\hline

\end{tabular}

\end{table*}

With available posts, replies and quotes/reshares from Gab, we construct conversation cascades, where each cascade is one complete conversation. Nodes in each cascade represent original post/reply/quote and edges represent reply/quote of post/reply/quote. The formal definition and the construction process of conversation cascades are given below:

\textbf{Cascade construction:} A conversation cascade is a directed graph $G_c=(V_c,E_c)$, where $V_c$ is a set of posts/replies/quotes and $E_c$ is a set of edges connecting posts ordered by time. We represent a node/post in the cascade as $P(v,t)$, where $v \in V_c$ is a post appearing in the social media at time $t$. There exists a directed edge($v^{\prime},v$), when $P(v,t)$ receives a reply/quote $P(v^{\prime},t^{\prime})$, where $t^{\prime} > t$. This process continues each time when a user replies/quotes a post. 

Thus, root node in conversation cascades represents original post and replies and quotes take branches from the root node(original post). Nodes in second or above level in the cascade take branches again if such nodes in turn get any replies or quotes. In total we constructed \textbf{1,721,441} cascades from the Gab data, which comprise of \textbf{19,220,059} nodes/posts contributed by \textbf{173,581} users and \textbf{15,476,852} edges. 

We evidence from these cascades that conversations in Gab follow one of the five patterns as represented in Figure~\ref{fig:cascade_type}. These conversation cascade types give a generic representation of user engagement patterns over time for a post. These cascades are also used to represent linear and non-linear conversation patterns in Gab. Figure~\ref{fig:cascade_freq} gives number of cascades in each conversation cascade pattern. In Table~\ref{tab:cascade_statistics}, we report statistics of cascade depth, volume, number of unique user participation, and structural virality of all cascade types. We calculate the structural virality of a cascade of size $n$ using Wiener Index(WI)~\cite{goel2015structural} given in the Equation~\ref{eq:wiener_index}, where $d_{ij}$ is the shortest distance of nodes $i$ and $j$. 

\begin{equation}
\label{eq:wiener_index}
	WI = \frac{1}{n(n-1) \sum_{i=1}^{n} \sum_{j=1}^{n} d_{ij}}
\end{equation}

\begin{figure}[!ht]
	\centering
		\subfloat[Depth Distribution of cascades\label{fig:depth_dist}]{\includegraphics[scale=0.45]{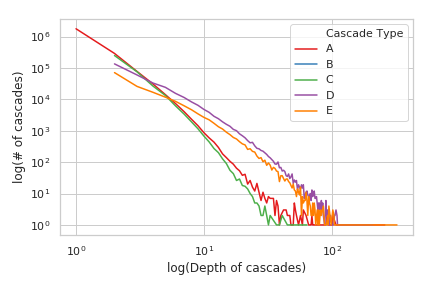}}	
	
		\hfill	
	
		\subfloat[Size Distribution of cascades\label{fig:size_dist}]{\includegraphics[scale=0.45]{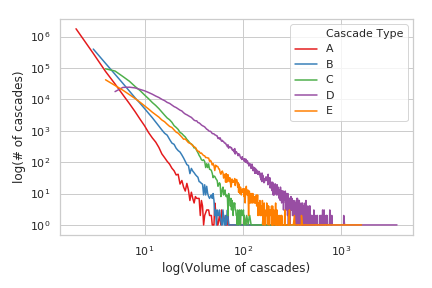}}

		\caption{Depth and Size distribution of each cascade type}
		\label{fig:depth_size}
\end{figure}

From Table~\ref{tab:cascade_statistics} and Figure~\ref{fig:depth_size}, we find that cascades of category E attains higher volume and depth with very few user participation and it plays a vital role in Gab conversations.

We also analyze depth and volume (number of nodes) of each cascade type and give their corresponding distributions in Figures~\ref{fig:depth_dist} and~\ref{fig:size_dist} respectively. From Figure~\ref{fig:depth_dist} we note that users in Gab have longer conversations, which take more branches after level 2 in the conversation thread(Cascade types D and E). Cascade type B is not shown in the Figure~\ref{fig:depth_dist} because these conversations split at the root node(post) and terminate at immediate next level. In Figure~\ref{fig:size_dist} we show distribution of volume(number of nodes/replies/quotes) across cascade types. The volume distribution of the given cascades are similar to the depth distribution given in Figure~\ref{fig:depth_dist}(Cascade types D and E have much engaging participation). Interestingly, we find that cascades of type B have more participation, given that these cascades stop at level 1, compared to cascades of type A.

\subsection{Cascades across topics}
Different topics spread in different ways over the networks in terms of speed, number of participants and the dynamics of their cascades. We analyze how this intuitively accepted notion applies to cascades on Gab and what topics in particular differ significantly from the others. In section ~\ref{sec:topic_discovery} we introduce a procedure by which we classify hashtags into different topics. We give a representative set of hashtags and their topics in Table~\ref{tab:topics}. Then we use the labeled hashtags to classify the cascades where those hashtags appear. Our analysis shows that cascades on more controversial topics have different characteristics than other cascades. They tend to result in larger cascades, majority of them being the same cascade type (Type D). Cascades in these polarizing topics are generated by users that are more strongly connected, i. e. have higher {\em tie strength}. We specifically identified three topics, ``Antisemitism", ``Anti Islam", and ``White Supremacy" to be noticeably different from other topics in regards to the nature of the cascades in which they are discussed. 

\begin{figure}
\centering
\includegraphics[scale=0.45]{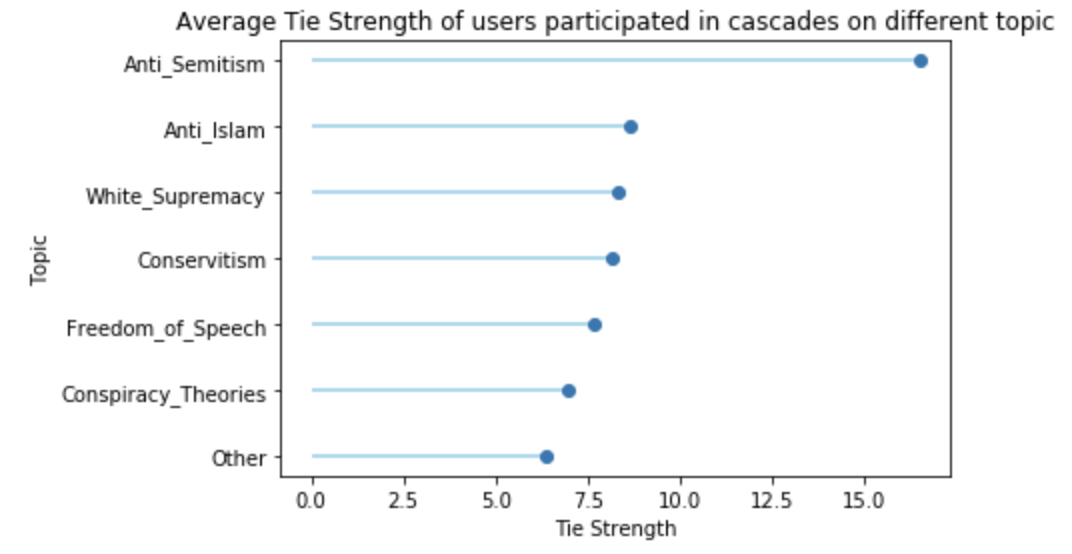}
\caption{Average tie strength of users participated in cascades on different topic. Users participating in cascades on polarizing topics like antisemitism, anti Islam, and white supremacy tend to have higher {\em tie strength}, i. e. the frequency of interactions between two users.} \label{fig:tiestrength}
\end{figure}

We define {\em tie strength} of user $u_1$ and $u_2$ as the number of times $u_1$ replies to a post from $u_2$ or $u_2$ replies to a post from $u_1$. Figure ~\ref{fig:tiestrength} shows that users who participated in topics of ``Antisemitism", ``Anti Islam", and ``White Supremacy" have higher {\em tie strength} and are more strongly connected which supports the theory that more controversial topics are adopted by users with higher tie strength~\cite{romeroAcrossTopics}.

ADL\footnote{www.adl.org} believes that white supremacist, hateful, antisemitic bigotry are widespread on Gab. Our findings are aligned with this statement and other studies that argue antisemitism and white nationalist topics are openly expressed on Gab and have great similarities in terms of communities who adopt these topics~\cite{quantitaveApproachToAntisemitism}~\cite{sorosMyth}.

\begin{table}[!t]
\caption{Topic Categories, a set of examples of hashtags, the number of instances, and the average size of cascades on each category} \centering
\label{tab:topics}
\centering

\begin{tabular}{|p{1.5cm}||p{3.6cm}||p{0.9cm}||p{0.9cm}|}
\hline
\textbf{Topic} & \textbf{Examples of Hashtags} & \textbf{Number of Cascades} & \textbf{Avg. Cascades Size} \\
\hline
\textbf{Conservitism} & \raggedright \#capitalism, \#conservitive \#freemarketmedicine, \#nationalsocialism, \#freedom & 7337 & 15.26 \\
\hline
\textbf{Anti Semitism} & \raggedright \#kikes, \#holocaust, \#nazis, \#jewsdid911, \#110neveragain, \#wearenotsemites, \#jewproof & 161 & 47.58\\
\hline
\textbf{White Supremacy} & \raggedright \#altright, \#whitegenocide, \#folkright, \#newright, \#retribalize \#whitetribalism & 7934 & 19.64 \\
\hline
\textbf{Freedom of Speech} & \raggedright \#speakfreely, \#1sta, \#censors, \#freespeech, \#censorship, \#shallnotcensor, \#freetommy & 13508 & 14.10 \\
\hline
\textbf{Anti Islam} & \raggedright \#banislam, \#bansharia, \#shariakills & 5379 & 16.01 \\
\hline
\textbf{Conspiracy Theories} & \raggedright \#qanon, \#pizzagate, \#q, \#wwg1wga, \#deepstate, \#thestorm, \#latearth & 23921 & 15.30 \\
\hline
\end{tabular}
\end{table}

\section{Topic Discovery of Cascades}
\label{sec:topic_discovery}
We introduce a novel procedure for labeling hashtags with the topics they belong to. We first looked at the top 200 most used hashtags on Gab. These hashtags then were classified by subject matter experts into 6 topics. Unlike other works identified on Twitter~\cite{romeroAcrossTopics}, our work do not cover a broad range of topics. Majority of posts on Gab are on political and rather controversial topics, thus we went few steps deeper and classified these political topics into more specific topics. We used tagdef\footnote{www.tagdef.com}, tagsfinder\footnote{www.tagsfinder.com}, Google, and Twitter to find the meaning of hashtags and assigned them to one of the 6 categories. Excluding hashtags that are too broad to be assigned to a specific topic, for instance \#gab, \#eu, \#music, or \#welcome, we labeled 126 hashtags.  

In the next step, we used our algorithm to label other hashtags used on Gab based on the 126 hashtags that we manually labeled. As shown in Algorithm~\ref{alg:cluster_hashtag}, the inputs of the method are the network of hashtags, list of topics with the set of hashtags assigned to each topic, and a constant value as threshold. We defined the network of hashtags as $G=\{V, E\}$ where $V$ is the set of hashtags used on Gab, and $E$ is the set of weighted edges. An edge between two vertices indicates that the two hashtags have appeared in the same post at least once and the weight of the edge represents the number of co-occurrences. We create this network of hashtags~\cite{7412726}, but we added weights to the edges to underscore the importance of the number of co-occurrences between two hashtags. In line 3 of the algorithm, an edge is passed to {\em get\_node\_with\_no\_topic()} method which returns the vertex that is not already assigned to any topic. The method returns null if both vertices of the edge are already assigned topics or if neither has any topic assigned to it. Because first, we are not interested in labeling hashtags that are already labeled and secondly, we cannot label a node if none of its neighbors is labeled. If the method returns a node $u$ we get the topics that are assigned to that node in step 6, and then for each topic $t$ in {\em topics}, we increment an integer property of node $v$, the other vertex of the edge, that represents $t$ by $w$, the weight of edge $e$. After all edges have been traversed and the properties of their respective vertices are updated we move to steps 9 to 13 where for each node $n$ in the graph and each topic $t$ in {\em topics}, we get the property $p_t$. If the value of the $p_t$ is greater than the threshold $\tau$, we conclude that node $n$, and the hashtag that it represents, belongs to topic $t$.

After we label hashtags with one or more topics, then for each cascade if hashtags of a topic $t_1$ appear more than C times, we assign that cascade to $t_1$. Note that a cascade could belong to more than one category. Table~\ref{tab:topics} shows the 6 topics we identified, some examples of the hashtags in each category, as well as the number of cascades and average size of cascades on each topic.

\begin{figure}[!ht]
	\centering	
		\subfloat[Figure a: Manually Labeled hashtags \label{fig:hashtag_cluster_actual}]{\includegraphics[scale=0.45]{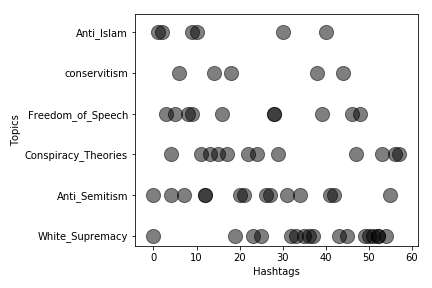}}	
	
		\qquad	
	
		\subfloat[Figure b: Prediction of our Algorithm \label{fig:hashtag_cluster_prediction}]{\includegraphics[scale=0.45]{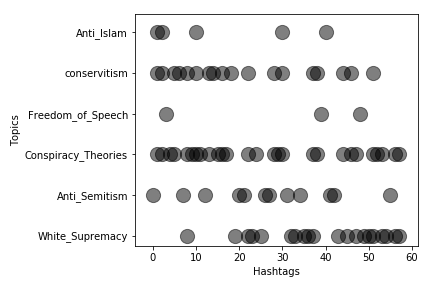}}

		\caption{Accuracy of the hashtag labeling algorithm, ~\ref{fig:hashtag_cluster_actual} shows the ground truth, i. e. manually labeled hashtags, and ~\ref{fig:hashtag_cluster_prediction} shows how our algorithm classified the same hashtags. Overall accuracy of the algorithm is 71\% as given in Table~\ref{tab:evaluation_result}}
		\label{fig:evaluation_hashtag_labeling}
\end{figure}

To evaluate our algorithm, we designed a procedure where we randomly picked 42 hashtags from the set of 126 labeled hashtags and labeled the rest of the hashtags in the set using our algorithm. Then we compared the results of our algorithm with our manual labeling. FIgure~\ref{fig:evaluation_hashtag_labeling} shows the distribution of hashtags among different topics. Figure~\ref{fig:hashtag_cluster_actual} shows the ground truth, i. e. manually labeled hashtags, and figure~\ref{fig:hashtag_cluster_prediction} shows the results of our algorithm.

Since one hashtag could belong to multiple topics, we used the evaluation metrics of multi-label learning algorithms mentioned in ~\cite{6471714}. Table~\ref{tab:evaluation_result} shows how well our algorithm performs in labeling the hashtags with topics, we use accuracy, recall, precision, $F1$ score, Hamming Loss (HL), and Subset Accuracy (SA).

\begin{table}[!t]
\caption{Performance of hashtag labeling algorithm. Our algorithm produced high results for conventional performance metrics of accuracy, recall, and precision. Low HL (Hamming Loss) and high SA (Subset Accuracy) are also strong indicatives that our algorithm performs well in classifying hashtags into topics.} \centering
\label{tab:evaluation_result}
\centering

\begin{tabular}{|p{0.8cm}||p{1cm}||p{0.7cm}||p{0.9cm}||p{0.4cm}||p{0.6cm}||p{0.6cm}|}
\hline
\textbf{Metric} & Accuracy & Recall & Precision & F1 & HL & SA \\
\hline
\textbf{Value} & 71\% & 73\% & 91\% & 77\% & 16\% & 84\% \\
\hline
\end{tabular}
\end{table}

\begin{algorithm}
\caption{Topic Discovery of Hashtags}\label{alg:cluster_hashtag}
\begin{algorithmic}[1]
\Require G, $\mathbf{T}$ and $\tau$
\Procedure{label\_hashtags}{}
\For{$e \in E$}
	\State \textit{u} = get\_node\_with\_no\_topics(e)
	\State \textit{v} = $e-u$
	
    \If{$u \neq \emptyset$}
    	\State $topics=$ get\_topics(\textit{v})
    	\For{$t \in topics$}
        	\State inc\_node\_prop(\textit{u},t,\textit{w}) 
        \EndFor
    \EndIf
 
\EndFor

\For{$n \in V$}
	\For{$t \in T$}
		$p_t =$ get\_property(G,n,t)
		\If{$p_t > \tau$}
    		\State add\_hashtag\_to\_topic(G,n,t)
    	\EndIf
	\EndFor
\EndFor

\EndProcedure
\end{algorithmic}
\end{algorithm}

\section{Cascade Analysis}
\label{sect:evolution}

\subsection{Response rate in cascades}
\begin{figure}
\centering
\includegraphics[scale=0.5]{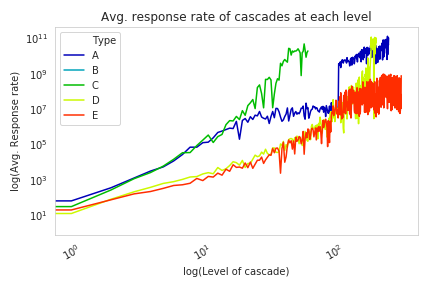}
\caption{Avg. response rate at each level/depth of the cascade. Higher values represent faster average response time. Simpler cascades(types A and C) have higher response rates than complex cascade structures(types D and E).} \label{fig:cascade_responses}
\end{figure}

With our proposed cascade types in Section~\ref{sect:cascades}, we analyze response time of each cascade type. With this experiment, we aim to produce results that depict how fast posts in these cascades get responses and help growing/evolving the cascades. We also provide a notion for response rate to compare velocity of responses at all levels in a given cascade type. We define response rate($\mathcal{R}_c$) of a cascade type as the average speed of response(s) of posts at a given level/depth of a cascade type($c$). Equation~\ref{eq:response_rate} gives a formulation to calculate response rate at a given level/depth ($l$) for a given cascade ($c$) and $\Delta_l^c$ is an average response time for posts at a given level/depth ($l$) for a given cascade type ($c$).

\begin{equation}
	\label{eq:response_rate}
	\mathcal{R}_c^l = \frac{1}{\Delta_c^l}
\end{equation}

\begin{figure}
\centering
\includegraphics[scale=0.5]{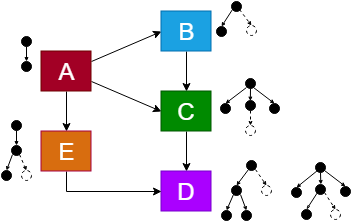}
\caption{Overview of cascade evolution as a state diagram. All cascades start as type A and they can evolve to the maximum of type D.} \label{fig:cascade_evolution}
\end{figure}

\begin{figure}[!ht]
	\centering	
		\subfloat[Figure a: Time taken by a cascade type to evolve into another\label{fig:evolution_time}]{\includegraphics[scale=0.5]{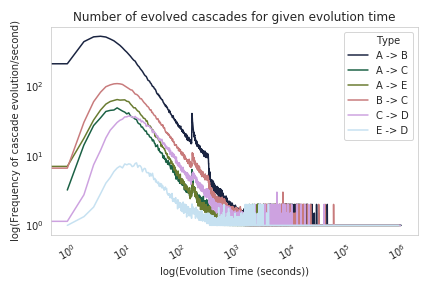}}	
	
		\qquad	
	
		\subfloat[Figure b: No. of post for a cascade type to evolve into another\label{fig:evolution_post}]{\includegraphics[scale=0.5]{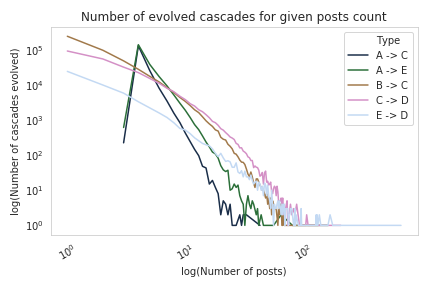}}

		\caption{Summary distributions of amount of time and posts taken by a cascade to evolve into another. Sudden spikes in the plot are due to anomalies in the data. 30\% of cascades in each cascade type require less than \textit{2 minutes} and the number of cascade evolution decreases as the number of posts increases}
		\label{fig:evolution_statistics}
\end{figure}

The distribution of average response rates of all cascade types at each level is given in Figure~\ref{fig:cascade_responses}. Interestingly, we find that all cascade types start with almost equal and slower response rate and progress with faster responses over time. Importantly, cascades of type \textbf{C} achieve overall higher response rate at earlier levels, even though they do not grow as larger and deeper as other cascades. Also, cascades of type \textbf{A} follow constant response rate like other cascade types and spikes as depth and volume increase. We also find that response rate for other larger cascades such as type \textbf{D} and \textbf{E} is inversely proportional to the distribution of volume of the cascade. 

\subsection{Evolution of cascades}

\begin{figure}
\centering
\includegraphics[scale=0.55]{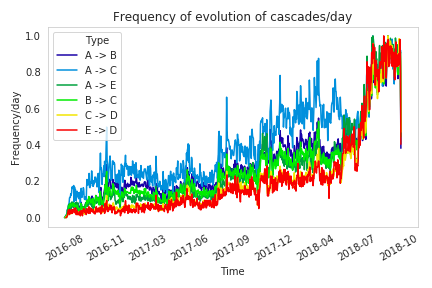}
\caption{Normalized timeseries distribution of number of cascade evolution for each evolution type.} \label{fig:evolution_timeseries}
\end{figure}

Given cascade types of Gab conversations, we study their growth patterns and evolution. All Gab conversations/cascading patterns, as given in Figure~\ref{fig:cascadetype}, starts with type \textbf{A} and some of them moderately evolve into other cascade types. All possible transformations within the proposed cascade varieties are given in Figure~\ref{fig:cascade_evolution}. It is notable from this figure that a conversation cascade can reach its maximum potential by transforming to cascade type \textbf{D} and cascades must evolve into other types(B,C,E) before reaching type \textbf{D}. Providing such evolution patterns and number of occurrences of each cascade from Figure~\ref{fig:cascade_freq}, we find that there are significant evolution of cascades in our data.

With the availability of evolution patterns in Gab conversations, we provide basic analysis such as time~\ref{fig:evolution_time} and number of posts~\ref{fig:evolution_post} required by a cascade to evolve into another. As a summary of this plot, we present Table~\ref{tab:evolution_statistics} to give minimum, maximum, average, and standard deviation of number of time and posts required to achieve evolve the cascades.

\begin{table}[!t]
\caption{Number of posts and time required by a cascade type to evolve into another. Overall evolution in Gab is slow with smaller number of posts and longer time to perform evolution.} \centering
\label{tab:evolution_statistics}
\centering

\begin{tabular}{|p{0.8cm}||p{0.6cm}||p{0.6cm}||p{0.6cm}||p{0.6cm}||p{0.6cm}||p{0.6cm}||p{0.6cm}|}
\hline
\textbf{Evolution Type} & Measure & $A \rightarrow B$  & $A \rightarrow C$ & $A \rightarrow E$ & $B \rightarrow C$ & $C \rightarrow D$ & $E \rightarrow D$ \\
\hline
\multirow{4}{*}{\shortstack[l]}{\textbf{\# replies to evolve}} & Min. & 2 & 3 & 3 & 1 & 1 & 1 \\\cline{2-8}
 & Max. & 3 & 43 & 163 & 134 & 208 & 675 \\\cline{2-8}
 & Avg. & 3 & 4.4 & 4.85 & 2.59 & 3.91 & 3.58 \\\cline{2-8}
 & Std. Dev. & 0.03 & 1.03 & 1.92 & 3.11 & 4.95 & 7.53 \\
\hline
\hline
\multirow{3}{*}{\shortstack[l]}{\textbf{Time to evolve (hrs)}} & Min. & 5s & 32s & 10.5s & 11.9s & 26s & 63s \\\cline{2-7}
 & Max. & 18742 & 18625 & 17069 & 17838 & 18421 & 17036 \\\cline{2-7}
 & Avg. & 30.41 & 40 & 38.02 & 38.69 & 55.98 & 52.86 \\\cline{2-7}
\hline
\end{tabular}
\end{table}

Figure~\ref{fig:evolution_timeseries} gives a timeseries on all possible cascade evolution in Gab. Modeling this evolution helps to study intensification of conversations in Gab. Conversations in any social media intensifies when it create interests or controversies among users about the topic. We use traditional models, like Susceptible-Infected~\cite{banks2013growth} and Bass~\cite{mahajan1990new} models to fit evolution patterns that exist in Gab conversation cascades. We prefer to use these models, because of their ability to fit sigmoid curves which in turn maps exponential growth or exponential fall~\cite{zang2016beyond}.

\begin{figure*}[!ht]
	\subfloat[$A \rightarrow B$]{%
    	\includegraphics[width=0.3\textwidth]{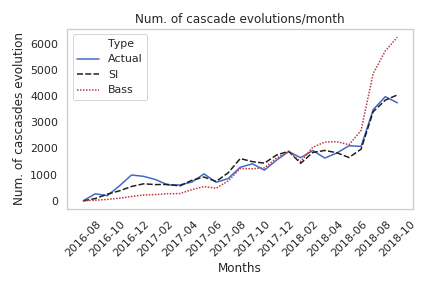}
    	\label{fig:evolution_atob}
	}
    \hfill
    \subfloat[$A \rightarrow C$]{%
       \includegraphics[width=0.3\textwidth]{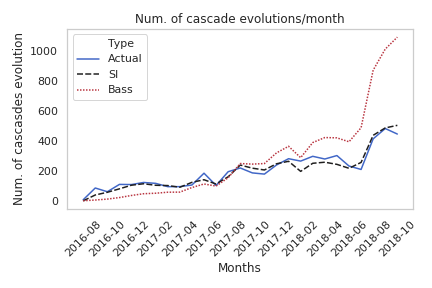}
       \label{fig:evolution_atoc}
    }
    \hfill
    \subfloat[$A \rightarrow E$]{%
       \includegraphics[width=0.3\textwidth]{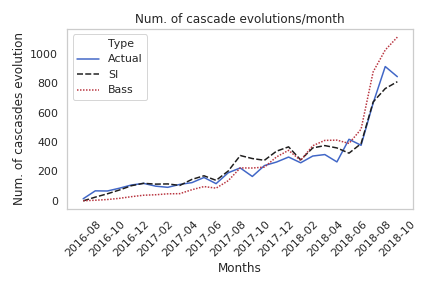}
       \label{fig:evolution_atoe}
    }
    
    \qquad
    
    \subfloat[$B \rightarrow C$]{%
    	\includegraphics[width=0.3\textwidth]{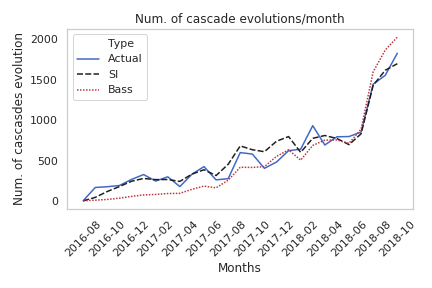}
    	\label{fig:evolution_btoc}
	}
    \hfill
    \subfloat[$C \rightarrow D$]{%
       \includegraphics[width=0.3\textwidth]{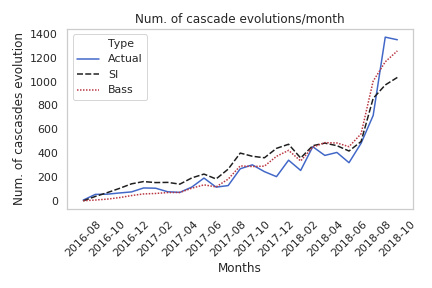}
       \label{fig:evolution_ctod}
    }
    \hfill
    \subfloat[$E \rightarrow D$]{%
       \includegraphics[width=0.3\textwidth]{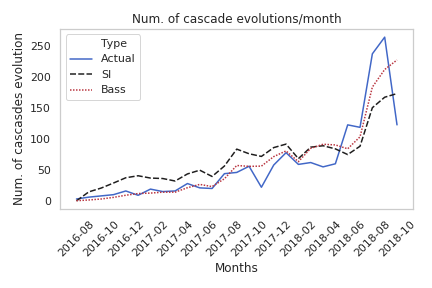}
       \label{fig:evolution_etod}
    }
    \caption{SI and Bass model fit for multiple cascade evolution timeseries. SI model fits for simple evolution types, while the Bass model captures the sharp spike that occurs near the end of the timeseries.}
    \label{fig:evolution_models}
\end{figure*}

The goal of our models is to predict a number of new evolution($\frac{dn(t)}{dt}$) given a time $t$ and a cumulative sum of user parameters($\epsilon$). Equation~\ref{eq:si_model} gives a modified equation of SI model. We model $\alpha,\gamma < 1$ to restrict complete participation of susceptibles ($N-n(t)$) and infected ones($n(t)$) because of an assumption that not all of the previous evolution are responsible for the current evolution.
\begin{equation}
	\label{eq:si_model}
	(SI)\frac{dn(t)}{dt}= \epsilon\ *\ \beta\ *\ n(t)^{\alpha}\ *\ (N-n(t))^{\gamma}
\end{equation}
	where,\\ 
		\qquad $\beta$ is evolution rate of cascades\\
		\hspace{20 cm} $N$ is total number of the cascades that evolved\\
		$n(t)$ is the cumulative sum of cascades evolved at time $t$. In other words, total infected ones at time $t$ 

\begin{figure}
\centering
\includegraphics[scale=0.45]{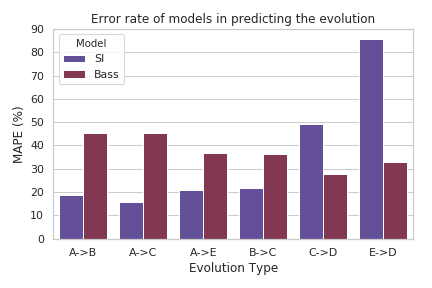}
\caption{Error(\%) of SI and Bass model to predict the evolution patterns. Our best model fit is 84\% accuracy(in $A \rightarrow C$ cascade evolution)} \label{fig:model_error}
\end{figure}

In Equation~\ref{eq:bass_model}, we give the Bass model with the user parameter($\epsilon$). Although, Bass model is introduced to describe the process of how new business product is taking effect in population, the model has been widely used in understanding diffusion and influence patterns in social networks also. Like the SI model, Bass model also generates S-shaped curve to fit exponential growth patterns.
\begin{equation}
	\label{eq:bass_model}
	(Bass)\frac{dn(t)}{dt}= \epsilon\ *\ m *\ \frac{(p+q)^2}{p} \frac{e^{-(p+q)t}}{(1+\frac{p}{q}e^{-(p+q)})^2}
\end{equation}
	where, 
		$m$ is total number of potential adopters \\
		$p$ is the parameter to model external influences\\
		$q$ is the parameter to model internal influences

Results of both models to map the cascade evolution is given in Figure~\ref{fig:evolution_models}. Each plot in this figure represents their corresponding evolution type, for example Figures~\ref{fig:evolution_atob} and~\ref{fig:evolution_btoc} marks result for the evolution types $A \rightarrow B$ and $B \rightarrow C$ respectively. From these results, we note that the performance of the SI model degrades as the evolution types become complex(for example, types $C \rightarrow$ and $E \rightarrow D$). We evaluate both models performance by the \textit{M}ean \textit{A}bsolute \textit{P}ercentage \textit{E}rror (\textit{MAPE}). Error rate of our models are given in Figure~\ref{fig:model_error}. Although this evolution problem itself can have its own model, we leave that work to focus in the future.

\section{Conclusion and Discussion}
Online extremism has gained momentum in the past decade due to extensive usage of social media. In this work we have given an extensive study on Gab using its conversations patterns and their related topics. We provide cascade templates for user conversations in Gab as conversation cascades. Dissecting Gab conversations as these cascade types give an intuition on analyzing more viral and responsive cascades. We provided variety of analysis that revolve around these cascades and given models that fits cascade evolution over time. Also, we studied about topics in the form of hashtag co-occurrence and given an algorithm to cluster hashtags into corresponding topics. 

In future, we plan to incorporate multiple social media forums like Gab, Twitter, and Reddit in the context of polarizing conversations and hate speech. We mainly focus to study information difussion and mutation patterns across online social media during shock events. Given this problem, there are various interesting areas to work in the near future. For example, we can engineer temporal features such as response rate, content features like word or sentence embedding, and features from ground truth network like follower network to model such information flow across platforms. We can embed these features in addition to post level features to predict the amount of hate in a social media. 

\bibliographystyle{IEEEtran} 
{\footnotesize
\bibliography{bibliography}}


\end{document}